\documentclass[twocolumn,epsfig]{revtex4}
\usepackage{epsfig}

\def\rhovec{\mbox{\boldmath $\rho$}}

\newcommand{\beq}{\begin{eqnarray}}
\newcommand{\eeq}{\end{eqnarray}}

\begin{document}

\title{Five-body cluster structure of 
double-$\Lambda$ hypernucleus $^{11}_{\Lambda \Lambda}$Be}

\author{E.\ Hiyama$^1$, M. Kamimura$^2$, Y. Yamamoto$^3$, and T. Motoba$^4$}

\address{$^1$RIKEN Nishina Center, 
RIKEN, Hirosawa 2-1,Wako,Saitama,
351-0198, Japan}

\address{$^2$Department of Physics, Kyushu University,
Fukuoka 812-8581, Japan}

\address{$^3$Physics Section, Tsuru University, Tsuru 402-8555, Japan}

\address{$^4$Laboratory of Physics, Osaka Electro-Comm.
University, Neyagawa 572-8530, Japan}

%

\begin{abstract}
Energy levels of the double $\Lambda$ hypernucleus, $^{11}_{\Lambda \Lambda}$Be
are calculated within the framework of an $\alpha \alpha n \Lambda \Lambda$
five-body model.
Interactions between constituent particles are determined
so as to reproduce reasonably the observed low-energy properties
of the $\alpha \alpha$, $\alpha \alpha n$ nuclei and the
existing data for $\Lambda$-binding energies of the
$\alpha \Lambda$, $\alpha \alpha \Lambda$, $\alpha n \Lambda$ and
$\alpha \alpha n \Lambda$ systems.
An effective $\Lambda \Lambda$ interaction
is constructed so as to reproduce, within the
$\alpha \Lambda \Lambda$ three-body model,
the $B_{\Lambda \Lambda}$ of $^6_{\Lambda \Lambda}$He,
which was extracted 
from the emulsion experiment, the NAGARA event.
With no adjustable parameters for
the $\alpha \alpha n \Lambda \Lambda$
system, $B_{\Lambda \Lambda}$ of the ground and
bound excited states of $^{11}_{\Lambda \Lambda}$Be
are calculated with the
Gaussian Expansion Method.
The Hida event, recently observed at KEK-E373 experiment,
is interpreted as an observation
of the ground state of the $^{11}_{\Lambda \Lambda}$Be.
\end{abstract}

\maketitle

\parindent 15 pt

In  nuclear physics involving strangeness, the fundamental problem is to describe the different facets of the interactions among the baryon octet in a unified way.
Our aim is to reveal various features of hyperon($Y$)-nucleon($N$) and $YY$ interactions through combined analyses for two- and many-body hyperonic systems.
For instance, $\Lambda N$ interaction models have been constructed so far by utilizing various $\Lambda$ hypernuclear data to complement the limited $\Lambda N$ scattering data.
The $\Lambda \Lambda$ interaction is an important entry into the $S\!=\!-2$ baryon-baryon interactions, decisive information about which is obtained from observations of double-$\Lambda$ hypernuclei and their separation energies for two $\Lambda$'s separated from a double-$\Lambda$ hypernucleus, denoted as $B_{\Lambda \Lambda}$.

In the KEK-E176/E373 hybrid emulsion experiments, there were observed several events corresponding to double-$\Lambda$ hypernuclei.
Among them was the  
epoch-making observation of the NAGARA event, 
which was identified uniquely as $^{\ 6}_{\Lambda \Lambda}$He 
 in the ground state 
 with a precise value of $B_{\Lambda \Lambda}=6.91 \pm 0.16$ MeV~\cite{Nagara,Nakazawa09}.
A second important observation was the Demachi-Yanagi event~\cite{DY,Nakazawa09} 
identified as $_{\Lambda \Lambda}^{10}$Be 
with $B_{\Lambda \Lambda}= 11.90 \pm 0.13 $ MeV\footnote{The    
 $B_{\Lambda \Lambda}$ values of NAGARA and Demachi-Yamanagi events
   are given as $7.25 \pm 0.19$ MeV and $12.33^{+0.35}_{-0.21}$ MeV 
in \cite{Nagara,DY} and \cite{DY}, respectively.
 In Ref.\cite{Nakazawa09}, however, the E176/373 emulsion 
 events have  been reanalyzed by using the new $\Xi^-$ mass \cite{PDG} .
 Their newly-obtained values of $B_{\Lambda \Lambda}$ for 
  these events are shown in the text.}                     
though it was uncertain whether
this event was interpreted to be the ground state or an excited state.

A newly observed double-$\Lambda$ event 
has been recently reported, called  the Hida event \cite{Nakazawa09}.
This event has two possible interpretations:
One is $^{11}_{\Lambda \Lambda}{\rm Be}$ with
$B_{\Lambda \Lambda}=20.83 \pm 1.27$ MeV, and 
the other is $^{12}_{\Lambda \Lambda}{\rm Be}$ 
with $B_{\Lambda \Lambda}=22.48 \pm 1.21$ MeV.
It is uncertain whether this is an observation
of a ground state or an excited state.

In the planned experiments at J-PARC, 
dozens of emulsion events for double-$\Lambda$ hypernuclei will be produced.
In emulsion experiments, however, it is difficult to 
determine the spin-parity or even to know whether an observed event 
corresponds to a ground or excited state. Therefore, it is vitally important
to compare the emulsion data with theoretical analyses to obtain a proper interpretation. 
%

In order to interpret the 'Demachi-Yanagi' event,
we studied in Ref.\cite{DY} the $^{10}_{\Lambda \Lambda}$Be hypernucleus
within the framework of the $\alpha \alpha \Lambda \Lambda$ 
four-body cluster model,
where the $\Lambda \Lambda$ interaction was taken consistently with
the NAGARA event.
Our calculated value for the $2^+$ state was  in good agreement with
the observed data.
Thus, the Demachi-Yanagi event is
interpreted as an observation of the $2^+$ excited state
of $^{10}_{\Lambda \Lambda}$Be.

The aim of this paper is to interpret the new Hida event 
on the basis of our theoretical study, adapting the
method used for interpreting the Demachi-Yanagi event.
At  present, the
Hida event has two possible interpretations:
$^{11}_{\Lambda \Lambda}$Be and $^{12}_{\Lambda \Lambda}$Be.
In this paper, we assume this event is
a $^{11}_{\Lambda \Lambda}$Be 
hypernucleus.
It is reasonable to employ 
an $\alpha \alpha n \Lambda \Lambda$ five-body model
for the study of $^{11}_{\Lambda \Lambda}$Be,
because, as mentioned above, the interpretation of 
the Demachi-Yanagi event for $^{10}_{\Lambda \Lambda}$Be
was possible on the basis of an
$\alpha \alpha \Lambda \Lambda$ four-body cluster model,
and $^{11}_{\Lambda \Lambda}$Be is 
composed of  $^{10}_{\Lambda \Lambda}$Be
plus one additional neutron.
We further note that the core nucleus $^9$Be is well described by
using an $\alpha \alpha n$ three-cluster model \cite{Arai}, 
and, therefore, it should be possible to model the structure change
of $^9$Be due to the addition of the two $\Lambda$ particles as a five-body problem.

The $\alpha \alpha n \Lambda \Lambda$
five-body cluster model employed in this paper is 
quite challenging as a numerical computation,
because of
the following conditions:
(1) there exist three species of particles 
($\alpha$, $\Lambda$, and neutron),
(2) five different kinds of interactions
($\Lambda$-$\Lambda$,
$\Lambda$-neutron, $\Lambda$-$\alpha$,
neutron-$\alpha$, and $\alpha$-$\alpha$) are involved,
(3) one must take into account the Pauli principle between
the two $\alpha$ particles and between the $\alpha$ and neutron.
%
We have succeeded in developing our Gaussian Expansion Method 
\cite{Hiyama03}, used in the above mentioned study 
of $^{10}_{\Lambda \Lambda}$Be, in order to perform
this five-body cluster-model calculation.

It should be emphasized, before going to the five-body calculation, that
all the interactions are determined so as to 
reproduce the observed binding energies of the two- and three-body
subsystems ($\alpha \alpha$, $\alpha n$, $\alpha \alpha n$, 
$\alpha \Lambda$, $\alpha \alpha \Lambda$,
$\alpha \Lambda \Lambda$, and $\alpha n\Lambda$).
We then calculated the energies of the $A=10$ four-body subsystems, 
$^{10}_\Lambda$Be
($=\alpha \alpha n \Lambda$) and  
$^{10}_{\Lambda \Lambda}$Be ($=\alpha \alpha \Lambda \Lambda$),
and found that they are 
simultaneously reproduced with no additional adjustable parameters needed.
We are then able to calculate both the ground and excited states
of $^{11}_{\Lambda \Lambda}$Be with no free parameter.
On the basis of careful calculations, we shall
show that the recently reported Hida event can be 
considered to be
an observation of the ground state of  
$^{11}_{\Lambda \Lambda}$Be.

\begin{figure}[htb]
\begin{center}
\epsfig{file=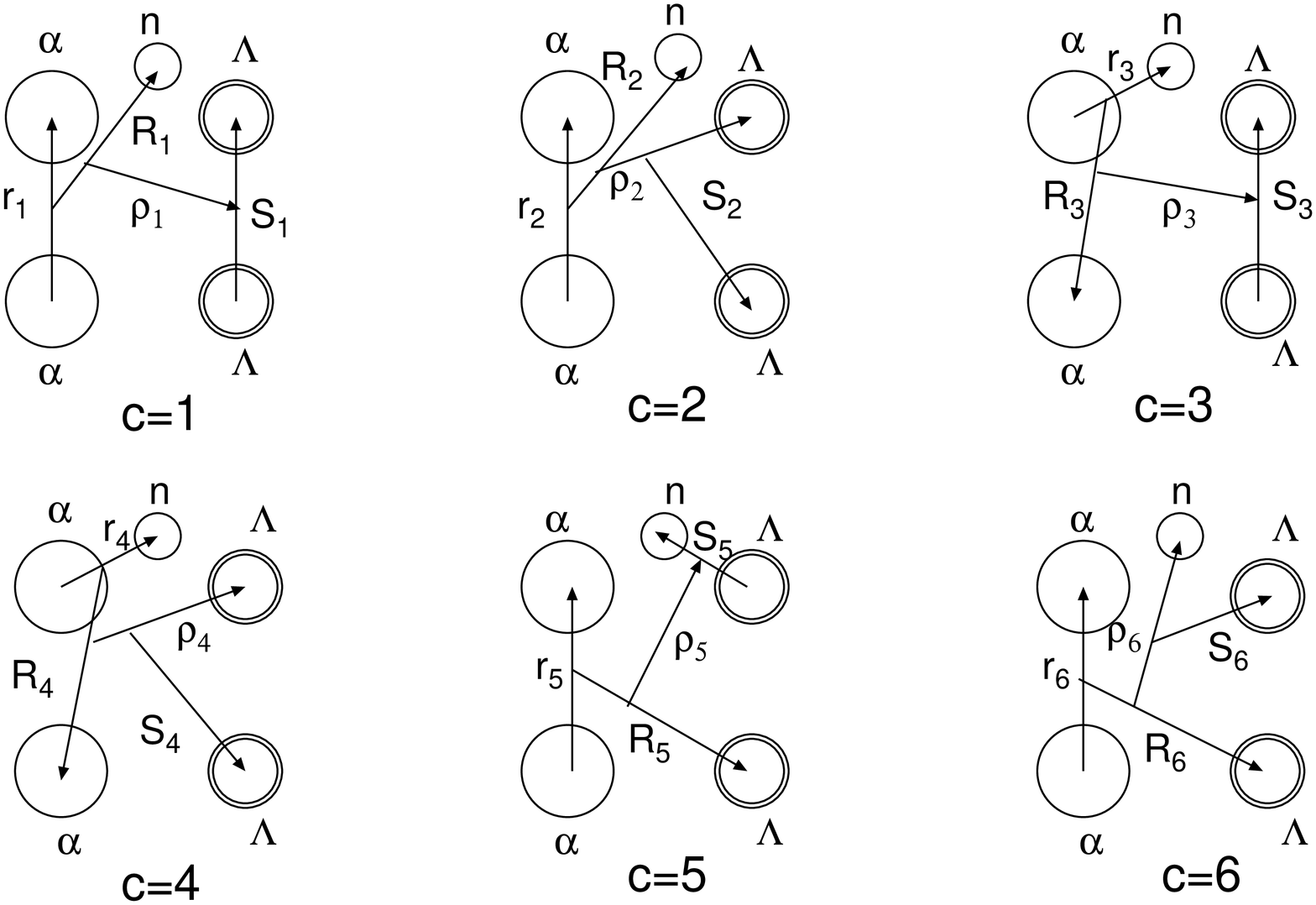,scale=0.25}
\end{center}
\caption{Example sets $(c=1-6)$ of Jacobi coordinates 
of the $\alpha  \alpha n \Lambda \Lambda$ five-body 
system. Antisymmetrization (symmetrization) 
of two  $\Lambda$'s (two $\alpha$'s) is to be made.
}
\label{fig:be11-jacobi}
\end{figure}

In order to take into account the full five-body degrees of freedom
of the $\alpha \alpha n \Lambda \Lambda$ system
and the full correlations among
all the constituent five particles,
we describe the total wave function,
$\Psi_{JM}(^{11}_{\Lambda \Lambda}{\rm Be})$,
as a function of the {\it entire} 35 sets of Jacobi coordinates 
$\{ {\bf r}_c, {\bf R}_c, \rhovec_c, {\bf S}_c ; c=1-35 \}$
in which the  two $\Lambda$'s (two $\alpha$'s) have been
antisymmetrized (symmetrized).
Some of the important coordinate sets $(c=1-6)$ are shown
in Fig.~\ref{fig:be11-jacobi}.

%

The total wave function is described
as a sum of five-body basis functions $ \Phi^{(c)}_{JM,\, \beta}$,
which is an extension of our previous work using
four-body basis functions \cite{Hiyama02}:
\begin{equation}
  \Psi_{JM}(^{11}_{\Lambda \Lambda}{\rm Be}) 
  = \sum_{c=1}^{35} \sum_\beta 
 C^{(c)}_\beta {\cal A}_{\Lambda \Lambda}{\cal S}_{\alpha \alpha}
  \, \Phi^{(c)}_{JM,\, \beta} \, ,
         \nonumber \\
\label{eq:totalwf}
\end{equation}
where ${\cal A}_{\Lambda \Lambda}$ (${\cal S}_{\alpha \alpha}$) 
is the antisymmetrizer (symmetrizer) for 
two $\Lambda$'s (two $\alpha$'s), and 
\begin{eqnarray}
&& \qquad \Phi^{(c)}_{JM,\, \beta}= 
    \xi(\alpha_1) \xi(\alpha_2) \nonumber \\
&& \times  \Bigg[ \, 
   \Big[\,\Big[\, [ \phi^{(c)}_{nl}({\bf r}_c)\,
         \psi^{(c)}_{NL}({\bf R}_c) ]_I \,
        \varphi^{(c)}_{n'l'}(\mbox{\boldmath $\rho$}_c) \Big]_K \,
         \Phi^{(c)}_{N'L'}({\bf S}_c)  \,
           \Big]_{L}  \nonumber \\
 \,\,&& \quad \times\,  \Big[\, [\chi_{\frac{1}{2}}(\Lambda_1)
              \chi_{\frac{1}{2}}(\Lambda_2)]_\Sigma \,
       \chi_{\frac{1}{2}}(n) \Big]_S  \,\,  
 \Bigg]_{JM} \, , 
\label{eq:5-bodybasis}
\end{eqnarray}
with $\beta \equiv  
 \{ n l , N L , n' l', N' L',\, IKL,\, \Sigma S \}$ 
denoting a set of the quantum numbers.
In Eq.~(\ref{eq:5-bodybasis}), $\xi(\alpha)$ is the internal
wave function of an $\alpha$-cluster having $(0s)^4$ configuration
and is used in the folding procedures for the $\alpha n, \alpha \Lambda$, and
$\alpha \alpha$ interactions. 
The $\chi_{\frac{1}{2}}(\Lambda)$ and  $\chi_{\frac{1}{2}}(n)$
are the spin functions of the $\Lambda$ and $n$, respectively.
Following Refs.~\cite{Hiyama02,Hiyama03}, 
the radial shapes of the basis function $\phi_{nlm}({\bf r})
(=r^l\:e^{-(r/r_n)^2}Y_{lm}({\widehat {\bf r}}))$  
are taken to be Gaussians 
with ranges postulated to lie in a geometric progression 
and similarly for 
$\psi_{NLM}({\bf R}), \varphi_{n'l'm'}(\mbox{\boldmath $\rho$})$
and  $\Phi_{N'L'M'}({\bf S})$.       
  The expansion coefficients $C^{(c)}_\beta$
and the eigenenergy $E$  of the 
total wave function   $\Psi_{JM}(^{11}_{\Lambda \Lambda}{\rm Be})$
are determined by solving the five-body Schr\"{o}dinger
equation using the Rayleigh-Ritz variational method.

In the present  $\alpha \alpha n \Lambda \Lambda$ five-body model for
$^{11}_{\Lambda \Lambda}$Be, it is absolutely necessary that any
subcluster systems composed of the two, three, or four constituent particles
are reasonably described by taking the interactions among these systems.
In our previous work  on double $\Lambda$ hypernuclei
with $A=7 -10$ within the framework
of the $\alpha x\Lambda \Lambda$ four-body 
cluster model ($x=n,p,d,t,^3\!{\rm He}$, and $\alpha$) \cite{Hiyama02}, 
the $\alpha$-$\alpha,
\alpha$-$n, \alpha$-$\Lambda, \Lambda$-$n$, 
and $\Lambda$-$\Lambda$ interactions
were determined so as to  reproduce
well the following observed quantities:
(i) Energies of the low-lying states and
scattering phase shifts in the $\alpha n$ and
$\alpha \alpha$ systems,
(ii) $\Lambda$-binding energies $B_{\Lambda}$ in
$^5_{\Lambda}$He ($=\alpha  \Lambda$), $^6_{\Lambda}$He
($=\alpha n \Lambda$) and
$^9_{\Lambda}$Be ($=\alpha  \alpha  \Lambda$),
(iii) double-$\Lambda$ binding energies  $B_{\Lambda \Lambda}$ in
$^6_{\Lambda \Lambda}$He $(=\alpha  \Lambda  \Lambda$),
the NAGARA event.
We then predicted, with no more adjustable parameters,
the energy level of $^{10}_{\Lambda \Lambda}$Be 
($=\alpha \alpha  \Lambda  \Lambda$), 
and found that, as mentioned 
before, the Demachi-Yanagi event was an observation of
the $2^+$ excited state of  $^{10}_{\Lambda \Lambda}$Be. 

In the  present five-body calculation, 
we employ the  interactions of Ref.~\cite{Hiyama02}
so that those severe constraints are also successfully met
in our two-, three-, and four-body subsystems.
But, the present core nucleus $^9$Be $(=\alpha \alpha n)$
does not belong to the subsystems studied previously. 
Since use of the same    
$\alpha \alpha$ and $\alpha n$ interactions  does not precisely
reproduce the energies of the low-lying states of $^9$Be
measured from the $\alpha \alpha n$ threshold 
(the same difficulty is seen in another 
microscopic $\alpha \alpha n$ cluster-model study
\cite{Arai}), we introduce an additional
phenomenological $\alpha \alpha n$ three-body force 
with a Gaussian shape, 
$v_0 e^{-(r_{\alpha-\alpha}/r_0)^2 - (R_{\alpha\alpha-n}/R_0)^2}$,
having $r_0=3.6$ fm, $R_0=2.0$ fm and 
$v_0=-9.7 \,{\rm MeV}\, (+13.0 \,{\rm MeV})$ for 
the negative-parity (positive-parity) state so as to
fit the observed energies of
the $3/2^-_1$ ground state and the $5/2^-_1, 1/2^-_1$ and $1/2^+_1$ 
excited states in $^9$Be. For the latter three resonance states, the same
bound-state approximation (namely, diagonalization of the Hamiltonian
with the $L^2$-integrable basis functions) was applied.  
Simultaneously, we found that the calculated
$B_{\Lambda}$ of 
the ground state $(1^-)$ of $^{10}_{\Lambda}{\rm Be}$
reproduced well the observed value.

In order to reproduce the observed 
$B_{\Lambda \Lambda}=6.91$ MeV for 
$^{\ 6}_{\Lambda \Lambda}{\rm He}$, we tuned
the $\Lambda \Lambda$ interaction \{Eq.(3.6) 
in Ref.~\cite{Hiyama02} \}
by multiplying the strength of the $i=3$ part by a factor 1.244.
 As for
$^{10}_{\Lambda \Lambda}{\rm Be}$, we obtained 
$B_{\Lambda \Lambda}^{\,{\rm cal}}(2^+_1)= 11.88$ MeV 
and $B_{\Lambda \Lambda}^{\,{\rm cal}}(0^+_1)= 14.74$ MeV, 
which  explain, respectively,
the  Demachi-Yanagi event for the
$2^+_1$ state ($11.90 \pm 0.13$ MeV) 
and the data of Ref.~\cite{Danysz63}
for the ground state ($14.6 \pm 0.4$ MeV; 
see Table 5 of Ref.\cite{Danysz63}).
The above successful check of the energies of the subsystems
encourages us to perform the five-body calculation
of $^{11}_{\Lambda \Lambda}{\rm Be}$ with
no adjustable parameter, expecting high reliability 
for the result.


\begin{figure}[htb]
\begin{center}
\epsfig{file=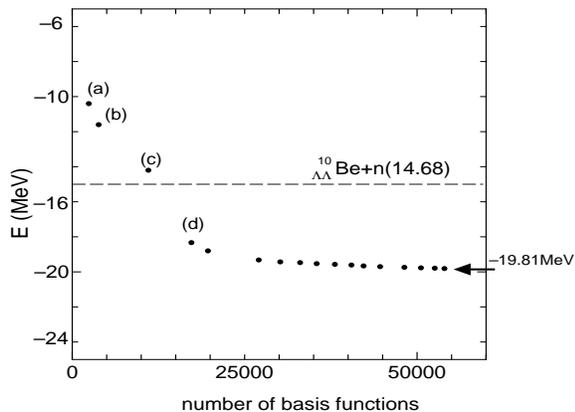,width=7.5cm,height=5.5cm}
\end{center}
\caption{Convergence of the energy $E$ of the $3/2^-$ ground state in
$^{11}_{\Lambda \Lambda}$Be with respect to increasing number of
the $\alpha \alpha  n \Lambda \Lambda $ five-body basis functions.
The energy is measured from
the five-body breakup threshold.
The dashed line shows the lowest threshold, $^{10}_{\Lambda \Lambda}$Be
$+n$ threshold.
See text for points (a)-(d).}
\label{fig:convergence}
\end{figure}

In Fig.~2, convergence of the calculated energy of the $3/2^-$
ground state of $^{11}_{\Lambda \Lambda}$Be is illustrated
with increasing number of five-body basis functions
$ \Phi^{(c)}_{JM,\, \beta}$ in  Eq.(1).
The energy is measured from the five-body breakup threshold.
The lowest threshold is $^{10}_{\Lambda \Lambda}{\rm Be} + n$, 
located at $-14.68$ MeV. 
The most important configuration in the ground-state wave function 
is found to be of the 
$^9{\rm Be}^* +\Lambda +\Lambda$ type, namely the configuration
described by using the Jacobi-coordinate sets  
$c=1-4$ in Fig.~1; 
here, $^9{\rm Be}^*$ denotes the 
$\alpha \alpha n$ three-body degrees of freedom. 
The energy point (a) is obtained by
taking $c=1$ configuration in Fig.1,
the point (b) is obtained by $c=1$ and $2$
and the point (c) is obtained by
$c=1$ to 4.
The additional $4$-MeV energy gain from point (c) to (d) is
obtained by including the configurations of
the $^9_\Lambda{\rm Be}^* +n +\Lambda$ type 
described with the Jacobi coordinates such 
as $c=5$ and $6$, in which  $^9_\Lambda{\rm Be}^*$ stands for
the $\alpha \alpha \Lambda$ degrees of freedom. 
Another $1.5$-MeV gain from point (d) down to
the converged value ($-19.81$ MeV with the accuracy of 10 keV
totally with $\sim 50,000$ basis functions)
is achieved by including all the other types of
configurations such as 
$^{10}_{\Lambda \Lambda}{\rm Be}^* +n $,
$^{10}_{\Lambda}{\rm Be}^* + \Lambda $,
$^{7}_{\Lambda \Lambda}{\rm He}^* + \alpha $,
$^6_\Lambda{\rm He}^* +~^5_\Lambda{\rm He}^*$.
Thus, the ground state is found to  be bound by 5.1 MeV
below the lowest threshold.
The angular momentum space of $l, l', L, L', \lambda \leq 2$
was found to be sufficient to obtain convergence for the
energy.


\begin{table}[h]
\caption{Calculated r.m.s. distances ${\bar r}_{\alpha-\alpha}$,
${\bar r}_{\alpha-n}$ 
and ${\bar r}_{(\alpha\alpha)-n}$ 
in $^9$Be,
$^{10}_{\Lambda}{\rm Be}$ and 
$^{11}_{\Lambda \Lambda}{\rm Be}$.}
\vskip 0.2cm
 \begin{tabular}{cccc}
 \hline
 \hline
(fm)  & ${\bar r}_{\alpha-\alpha}$ & ${\bar r}_{\alpha-n}$  
 & ${\bar r}_{(\alpha\alpha)-n}$  \\
\vspace {-4 mm} \\
\hline
\vspace {-4 mm} \\
$\quad$   $^{9}$Be $\quad$ &$\quad$ \, 3.68 $\quad$  &$\quad$  3.98$\quad$ 
 &$\quad$  3.54$\quad$  \\
   $^{10}_\Lambda$Be & $\quad$ 3.28$\quad$   & $\quad$ 3.53$\quad$  &
$\quad$  3.14$\quad$  \\
   $^{11}_{\Lambda \Lambda}$Be & $\quad$ 3.10$\quad$   &$\quad$\,  3.33
$\quad$  &$\quad$  2.94$\quad$  \\
\vspace {-3 mm} \\
 \hline
 \end{tabular}
\end{table}

It is interesting to look at the dynamical change of
the nuclear core, $^9$Be, which occurs due to the
addition of two $\Lambda$ particles.
The possibility of nuclear-core shrinkage due to a 
$\Lambda$-particle addition was originally pointed out
in Ref.~\cite{Motoba83} by  using the
$\alpha x \Lambda$ three-cluster model 
($x=n,p,d,t,^3\!{\rm He}$, and $\alpha$)
for $p$-shell $\Lambda$ hypernuclei.
As for the hypernucleus $^7_\Lambda$Li, the prediction of 
some 20\%-shrinkage, in Ref.~\cite{Motoba83} and in an updated 
calculation~\cite{Hiyama99},  was actually confirmed by 
experiment~\cite{Tanida01}.
As far as  $p$-shell double-$\Lambda$
hypernuclei are concerned, 
the present authors found \cite{Hiyama02}, using an
$\alpha x \Lambda \Lambda$ model,
that participation of the second $\Lambda$
particle can induce a further $\sim \!8$\% reduction in the distance 
between the $\alpha$ and $x$ in the nuclear core.
For the present five-body $\alpha \alpha n \Lambda \Lambda$ system,
 $^{11}_{\Lambda \Lambda}$Be, 
Table I shows the r.m.s. distances ${\bar r}_{\alpha-\alpha}$ 
between two $\alpha$ particles,
${\bar r}_{\alpha-n}$ between 
$\alpha$ and $n$ and 
${\bar r}_{(\alpha\alpha)-n}$ between 
$(\alpha \alpha)$ and $n$
in $^9$Be,
$^{10}_{\Lambda}{\rm Be}$ and 
$^{11}_{\Lambda \Lambda}{\rm Be}$.
When the first $\Lambda$ particle is added to $^9$Be, a 
reduction of those distances
is about 20\%, whereas
the adding a second $\Lambda$ reduces
the distances by about 6\% ; this is similar to the
case of double $\Lambda$ hypernuclei studied
with the $\alpha x \Lambda \Lambda$ four-body model \cite{Hiyama02}.

\begin{figure}[htb]
\begin{center}
\epsfig{file=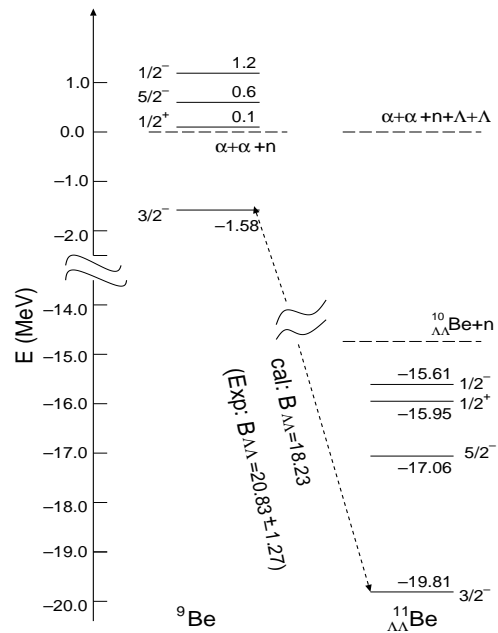,width=6.5cm,height=8.5cm}
\end{center}
\caption{Calculated energy spectra of the low-lying states of
$^{11}_{\Lambda \Lambda}$Be together with those of
the core nucleus $^9$Be.}
\label{fig:level}
\end{figure}

Finally, let us discuss the energy spectra of
$^{11}_{\Lambda \Lambda}$Be and its relation to the Hida event.
Using the same framework and interactions as those in
the $3/2^-$ ground state,
we calculated energies and wave functions of the
$5/2^-_1$, $1/2^+_1$ and $1/2^-_1$ states of  $^{11}_{\Lambda \Lambda}$Be;
there is no other bound state below the lowest 
$^{10}_{\Lambda \Lambda}{\rm Be}+n $ threshold.
The energy level is illustrated in  Fig.~\ref{fig:level} together with
that of $^9$Be.
Interestingly enough, the order of the $1/2^+$ and $5/2^-$ states is
reversed from $^9$Be to $^{11}_{\Lambda \Lambda}{\rm Be}$.
This is because the energy gain due to the addition of the
$\Lambda$-particle(s) is larger in the compactly coupled state 
($5/2^-_1)$ 
than in the loosely coupled state ($1/2^+_1$).
Note that the same type of 
theoretical prediction was reported, 
in our early work \cite{Hiyama00} for $^{13}_\Lambda$C
based on  the $\alpha \alpha \alpha \Lambda$ model, 
for the $\Lambda$ particle 
addition to the compactly coupled state ($3^-_1$)  
and  to the
loosely coupled state ($0^+_2$) in $^{12}$C.

As seen in Fig.~\ref{fig:level}
the calculated value of
 $B_{\Lambda \Lambda}(^{11}_{\Lambda \Lambda}$Be) is 18.23 MeV
for the $3/2^-$ ground state,
while for the excited states the  $B_{\Lambda \Lambda}$ values are
calculated to be less than 15.5 MeV.
Therefore, the observed Hida event can be
interpreted to be the ground state.
When our calculated binding energy is compared with
the experimental value of 20.83 MeV with a large uncertainty of 
$\sigma$=1.27 MeV,
we can say at least that our result does not contradict 
the data within 2$\sigma$.
%

So far, the $\Lambda \Lambda$ interaction strength has been
often estimated rather intuitively by the quantity
$\Delta B_{\Lambda \Lambda}(^{\rm A}_{\Lambda \Lambda}Z) \equiv
B_{\Lambda \Lambda}(^{\rm A}_{\Lambda \Lambda}Z)-2B_{\Lambda}
(^{\rm A-1}_{\Lambda}Z)$.
The observed value of $\Delta B_{\Lambda \Lambda}(^{\ 6}_{\Lambda \Lambda}$He)
is 0.67 MeV, to which our $\Lambda \Lambda$ interaction is adjusted.
The calculated value for the ground state of $^{10}_{\Lambda \Lambda}$Be
is 1.32 MeV. On the other hand, that for $^{11}_{\Lambda \Lambda}$Be
is obtained as only 0.29 MeV. Here, the $B_\Lambda(^{10}_\Lambda$Be) is
given by a weighted sum of the values for ground $1^-$ and excited $2^-$
states of $^{10}_\Lambda$Be so that there is no contribution of 
the $\Lambda N$ spin-spin interaction in the double-$\Lambda$ state.
One should notice here the remarkable difference between 
$\Delta B_{\Lambda \Lambda}$ values for $^{10}_{\Lambda \Lambda}$Be and 
$^{11}_{\Lambda \Lambda}$Be.

Furthermore, as discussed in the previous paper \cite{Hiyama02}, values of
$\Delta B_{\Lambda \Lambda}$ include rearrangement effects in
nuclear cores due to participation of $\Lambda$ hyperons.
Then, we showed that
the  $V_{\Lambda \Lambda}^{\rm bond}$ defined
by Eq.~(\ref{eq:V-bond}) was useful for
estimating the strength of the 
$\Lambda \Lambda$ interaction
\begin{equation}
V_{\Lambda \Lambda}^{\rm bond}(^{\rm A}_{\Lambda \Lambda}Z)
 \equiv B_{\Lambda \Lambda}(^{\rm A}_{\Lambda \Lambda}Z)
-B_{\Lambda \Lambda}(^{\rm A}_{\Lambda \Lambda}Z:V_{\Lambda
\Lambda}=0) \ ,
\label{eq:V-bond}
\end{equation}
where $B_{\Lambda \Lambda}(^{\rm A}_{\Lambda \Lambda}Z:V_{\Lambda
\Lambda}=0)$ denotes the $B_{\Lambda \Lambda}$ value calculated
by putting $V_{\Lambda \Lambda}=0$.
Table II lists the calculated values of $V_{\Lambda \Lambda}^{\rm bond}$ 
for $^{\ 6}_{\Lambda \Lambda}$He, $^{10}_{\Lambda \Lambda}$Be and 
$^{11}_{\Lambda \Lambda}$Be which are similar to each other.
Thus, the obtained $\Lambda \Lambda$ bond energy in
$^{11}_{\Lambda \Lambda}$Be turns out to be reasonable in spite
of the small value of $\Delta B_{\Lambda \Lambda}$.
\begin{table}[h]
\caption{ $\Lambda \Lambda$ bond energy  
$V_{\Lambda \Lambda}^{\rm bond}(^{\rm A}_{\Lambda \Lambda}Z)$
defined by Eq.~(\ref{eq:V-bond}).
}
\vskip 0.2cm
 \begin{tabular}{ccccc}
 \hline
 \hline
\vspace {-4 mm} \\
 & 
$^{\ 6}_{\Lambda \Lambda}$He & 
$^{10}_{\Lambda \Lambda}$Be & 
$^{11}_{\Lambda \Lambda}$Be & \\
\vspace {-4 mm} \\
 \hline
\vspace {-4 mm} \\
\vspace {-4 mm} \\
$V_{\Lambda \Lambda}^{\rm bond}(^{\rm A}_{\Lambda \Lambda}Z)$ 
 &$\quad$ \, 0.54 $\quad$  &$\quad$  0.53$\quad$ 
 &$\quad$  0.56$\quad$ &(MeV)$\quad$ \\
\vspace {-4 mm} \\
\vspace {-4 mm} \\
\hline
 \end{tabular}
\end{table}

In conclusion, 
motivated  by the recent observation of the Hida event
for a new double $\Lambda$ hypernucleus,
we have succeeded in performing a five-body calculation
of $^{11}_{\Lambda \Lambda}{\rm Be}$ using an
 $\alpha  \alpha  n \Lambda  \Lambda$ cluster model.
The calculated $\Lambda \Lambda$ binding energy does
not contradict the interpretation that the Hida event
is an observation of the ground state of $^{11}_{\Lambda \Lambda}{\rm Be}$. 
In this model, we described the core nucleus $^9$Be using the
$\alpha \alpha n$ three-cluster model and found a significant
structure change (shrinkage) of $^9$Be due to the addition of
the two $\Lambda$ particles.
For  an alternative interpretation of the Hida event 
as the ground (or any excited) state of $^{12}_{\Lambda \Lambda}$Be,
a corresponding six-body $\alpha \alpha nn \Lambda\Lambda$ model
calculation is necessary, but such an undertaking is 
beyond our present consideration. 
More precise data are needed in order to test our present 
result quantitatively.
In the near future, many data for double $\Lambda$ hypernuclei are 
expected to be found in the new emulsion experiment E07 at J-PARC. 
Then, our systematic predictions including the work of
Ref.~\cite{Hiyama02} will be clearly tested.
%

%
%
%

\section*{Acknowledgments}
The authors thank Professors K. Nakazawa, Th. A. Rijken
and B.F. Gibson for valuable discussions.
This work was supported by a Grant-in-Aid for
Scientific Research from Monbukagakusho of Japan.
The numerical calculations were performed on the 
HITACHI SR11000 at KEK
and FUJITSU PRIMEQUEST580 at Kyushu University.

\end{document}